\documentclass{ws-procs9x6}

\begin{document}

\title{Searching for sub-millisecond pulsars:\\
A theoretical view
\footnote{\uppercase{T}his work is supported by National
Scientific Foundation of China (10573002) and the Key Grant
Project of Chinese Ministry of Education (305001).}}

\author{R.~X. Xu (Xu/Renxin)}

\address{School of Physics, Peking University, Beijing 100871, China\\
E-mail: r.x.xu@pku.edu.cn}

\maketitle

\abstracts{%
Sub-millisecond pulsars should be triaxial (Jacobi ellipsoids),
which may not spin down to super-millisecond periods via
gravitation wave radiation during their lifetimes if they are
extremely low mass bare strange quark stars.
It is addressed that the spindown of sub-millisecond pulsars would
be torqued dominantly by gravitational wave radiation (with
braking index $n\simeq 5$).
The radio luminosity of sub-millisecond pulsars could be high
enough to be detected in advanced radio telescopes.
Sub-millisecond pulsars, if detected, should be very likely quark
stars with low masses and/or small equatorial ellipticities.
}%

\section{Introduction}

Historically, the very idea of ``{\em gigantic nucleus}'' (neutron
star) was first given by Landau more than 70 years ago, was soon
involved in supernova study of Baade and Zwicky, and seems to be
confirmed after the discovery of radio pulsars. Although this is a
beautiful story, no convincing work, either theoretical from first
principles or observational, has really proved this idea that
pulsars are normal neutron stars, since nucleon (neutron or
proton) was supposed to be point-like elementary particles in
Landau's time but not now.
An idea of quark stars, which are composed by free quarks rather
than free nucleus, was suggested as more and more sub-nucleon
phenomena were understood, especially the asymptotic freedom
nature of strong interaction between quarks; and it is found that
no problem exists in principle if pulsars are quark stars.
Therefore, astrophysicist without bias should think equally
neutron and quark stars to be two potential models for the nature
of pulsar-like stars\cite{xu06}.

One of the most important problems could then be differentiating
neutron and quark stars in the new millennium astrophysics.
Many likely ways to identifying a quark star are
proposed\cite{xu06}, but the search of sub-millisecond pulsars
would be an excellent key experiment.
The reason for this is very intuitive and almost
model-independent.
Normal neutron stars are gravitationally confined, and thus can
{\rm not} spin with periods being less than a break period of
$\sim 0.5M_1^{1/2}R_6^{-3/2}$ ms (mass: $M_1=M/M_\odot$, radius:
$R_6=R/10^6$ cm), but quark stars are chromatically confined,
without limitation by the Kepler frequency.
Additionally, a much high spin frequency may not damped
significantly if a quark star has a very low mass.
Therefore, sub-millisecond pulsars should be expected to be
detected if pulsars are (low-mass) quark stars, with a rapid spin
at birth, but could not exist if they are normal neutron stars.

We are viewing the implications of possibly future discovery of
sub-millisecond pulsars in this paper, with (solid) quark star
models to be focused for the nature of pulsar-like compact
objects.
The formation, evolution, and magnetospheric activity of
sub-millisecond pulsars are also discussed.
Because the pulse profiles of radio pulsars are highly modulated
($\sim 100\%$), with very small pulse-widths (a few $\sim 10^{\rm
o}$), the timing precision could only be high enough to uncover
sub-millisecond pulsars in radio band. We focus thus radio pulsars
here.

\section{Spindown of sub-millisecond pulsars}

\subsection{Spin periods limited by gravitational wave emission}

{\em Fluid pulsars}.
In Newtonian theory, a rapidly rotating fluid Maclaurin spheroid
is secularly unstable to become a Jacobi ellipsoid, which is
non-axisymmetric, if the ratio of the rotational kinetic energy to
the absolute value of the gravitational potential energy $T/|W|
> 0.1375$.
For a Maclaurin spheroid with homogenous density $\rho$, the ratio
$T/|W|=0.1375$ results in an eccentricity $e=0.81267$ (or
$c/a=\sqrt{1-e^2}\simeq 0.6$) and thus in a critical frequency
$\Omega_c\sim 5.6\times 10^3$/s (i.e., spin period $P_c\sim 1.1$
ms) if $\rho=4\times 10^{14}$ g/cm$^3$. Sub-millisecond pulsars
could then be Jacobi ellipsoids, to be triaxial.
In the general relativistic case\cite{chandra70,fs78},
gravitational radiation reaction amplifies an oscillation mode,
and it is then found that the critical value of $T/|W|$ for the
onset of the instability could be much smaller than 0.1375 for
neutron stars with mass of $\sim M_\odot$.
This sort of non-axisymmetric stellar oscillations will inevitably
result in gravitational wave radiation, and put limits on the spin
periods.

A kind of oscillation mode, socalled $r$-mode, is focused on in
the literatures\cite{and98,fm98,lom98}.
The $r$-mode oscillation is also called as the Rossby waves that
are observed in the Earth's ocean and atmosphere, the restoring
force of which is the Coriolis force.
This instability may increase forever if no dissipation occurs.
Therefore, whether the instability can appear and how much the
oscillation amplitude is depend on the interior structure of
pulsars, which is a tremendously complicated issue in supranuclear
physics.

The Kepler frequency of low-mass bare strange stars could be
approximately a constant,
\begin{equation}
\Omega_0=\sqrt{GM\over R^3}=1.1\times 10^4~~{\rm s^{-1}},
\end{equation}
where the average density is taken to be $\sim 4\times 10^{14}$
g/cm$^3$. Correspondingly, the spin period
$P_0=2\pi/\Omega_0\simeq 0.6$ ms $<P_c$.
It is found by Xu\cite{xu06ap} that the gravitational wave
emissivity of quark stars is mass-dependent. The $r$-mode
instability could not occur in fluid bare strange stars with radii
being smaller than $\sim 5$ km (or mass of a few $0.1M_\odot$)
unless these stars rotates faster than the break frequency (in
fact, a more effective gravitational wave emission mode occurs if
$P<\sim P_0$, see Eq.(\ref{lossrate})). These conclusions do not
change significantly in the reasonable parameter-space of bag
constant, strong coupling constant, and strange quark mass.

Some recent observations in X-ray astronomy could hint the
existence of low-mass bare strange stars\cite{xu05}.
The radiation radii (of, e.g., 1E 1207.4-5209 and RX J1856.5-3754)
are only a few kilometers (and even less than 1 km).
No gravitational wave emission could be detected from such fluid
stars even they spin only with a period of $\sim 1$ ms.

{\em Solid pulsars}.
A protoquark stars should be in a fluid state when their
temperatures are order of 10 MeV, but would be solidified as they
cool to very low temperatures\cite{xu03a,xu05a}.
Assuming the initial ellipticity of a solid quark star keeps the
same as that of the star just in its fluid phase, strain energy
has to develop when a solid quark star spins down.
Quake-induced glitches of the quark star occur when the strain
energy reaches a critical value\cite{z04}, and we thus suggest
that the stellar ellipticity would be approximately determined by
the conventional Maclaurin spheroids (for $P\gg 1$ ms)
\begin{equation}
\varepsilon(P)\simeq {5\Omega^2\over 8\pi G\rho}\simeq 3\times 10^{-3}P_{\rm 10ms}^{-2},%
\label{varepsilon}
\end{equation}
where the spin period $P=2\pi/\Omega=P_{\rm 10ms} \times 10$ ms,
provided that the density $\rho\simeq 4\times 10^{14}$ g/cm$^3$ is
a constant.

A pulsar must be non-axisymmetric in order to radiate
gravitationally. A triaxial pulsar, with deformation ellipticity
$\epsilon_e$ in its equatorial plane, or a wobbling pulsar, either
freely or forcedly, may thus radiate gravitational waves, the
frequency of which is $2\Omega$ for the former but is
$\Omega+\Omega_{\rm prec}$ (the precession frequency $\Omega_{\rm
prec}$ is orders of magnitude smaller than $\Omega$) for the
later.
This wave results in a perturbed metric, which is order of $h_0$
being given by\cite{kip95},
\begin{equation}
h_0={128\pi^3G\rho_0\over 15c^4}\cdot { R^5\over
dP^2}(\epsilon_e~{\rm or}~\epsilon\theta) \approx
2.8\times 10^{-20} R_6^5d_{\rm kpc}^{-1}P_{\rm 10ms}^{-2}(\epsilon_e~{\rm or}~\epsilon\theta),%
\label{h0}
\end{equation}
where approximations $I\simeq 0.4MR^2$ and $M\simeq 4\pi
R^3\rho_0/3$ are applied for solid quark stars in the right
equation, the pulsar's distance to earth is $d=d_{\rm kpc}\times
1$ kpc, $\theta$ is the wobble angle.

For normal neutron stars, $\epsilon_e$ and $\epsilon\theta$ are
supported by crustal shear stress and magnetic pressure.
However for solid quark stars, this mechanisms may not work due to
a relatively negligible magnetic and Coulomb forces.
Nevertheless, glitches of solid quark stars could also produce
bumps, with a maximum ellipticity\cite{owen05},
\begin{equation}
\epsilon_{\rm max} \sim 10^{-3} ({\sigma_{\rm max}\over
10^{-2}})R_6^{-6}(1+0.084R_6^2)^{-1},
\label{emax}
\end{equation}
where $\sigma_{\rm max}$ is the stellar break strain. This
ellipticity is larger for low-mass quark stars due to weaker
gravity. This maximum ellipticity could be much smaller than the
ellipticity of Maclaurin spheroids with $P<\sim 1$ ms
(sub-millisecond pulsars).
Therefore, for the sake of simplicity, we assume that the real
ellipticity of a solid quark star could be $\varepsilon(P)$, due
to stress releases through star-quake induced glitches\cite{z04},
in the discussion below.

LIGO is sensitive to hight frequency waves, which recently puts
upper limits on $h_0$ for 28 known pulsars through the second LIGO
science run\cite{LIGO2005}. The upper limits are order of
$10^{-24}$, which means approximately an limit of $\epsilon
R_6^5<10^{-4}$.
Only three normal pulsars are targeted; others are millisecond
pulsars.
The upper limits of masses and radii for millisecond pulsars are
constrained\cite{xu06ap} by the second LIGO science run.
Specially, the radius of the fastest rotating pulsar, PSR
B1937+21, could be smaller than $\sim 2$ km if its wobble angle
$\theta$ is between $1^{\rm o}$ and $10^{\rm o}$.

Energy of gravitational wave is not as instinctive as the
perturbed metric, $h_0$.
Nonetheless, for triaxial sub-millisecond pulsars, the luminosity
of gravitational waves radiation from a solid pulsar could be
obtained\cite{weinb72}, and the total rotation energy loss via
gravitational and magnetospheric (photons and particles) emission
is
\begin{equation}
-I\Omega{\dot\Omega}={32G\Omega^6I^2\epsilon_e^2\over
5c^5}+{2\over 3c^3}\mu_m^2M^2\Omega^4,
\label{lossrate}
\end{equation}
where $\mu_m$ is the magnetic momentum per unit mass.
It is worth mentioned that the work relevant to rapid rotating
Jacobi ellipsoids was done by many authors\cite{j1,j2,j3,j4,j5}.
The ratio of the term due to gravitational wave and that due to
magnetospheric activity in Eq.(\ref{lossrate}) is
\begin{equation}
f_r={192GR_{\rm eff}^4\epsilon_e^2\over 125
c^2\mu_m^2}\Omega^2\simeq 4.5\times 10^{11}R_{\rm
effkm}^4\epsilon_e^2\mu_{m-6}^{-2} P_{\rm ms}^{-2},
\label{ratio}
\end{equation}
where an effective radius $R_{\rm eff}=R_{\rm effkm}\times (1$ km)
is defined through $I=2MR_{\rm eff}^2/5$, $\mu_m=\mu_{m-6}\times
(10^{-6}$ G cm$^3$ g$^{-1})$, $P=2\pi/\Omega=P_{\rm ms}\times (1$
ms).
It is evident from Eq.(\ref{ratio}) that the braking torqued by
gravitational wave dominates the spindown of a sub-millisecond
pulsar unless $\epsilon_e\ll 1$ and/or $R_{\rm eff}\ll 1$ km.
The braking index for gravitational wave emission is $n\equiv
\Omega{\ddot \Omega}/{\dot \Omega}^2=5$ from Eq.(\ref{lossrate}),
and this feature of $n>3$ could be evidence for gravitational wave
in return.
Therefore, gravitational wave radiation alone could result in an
increase of the rotation period of sub-millisecond pulsars,
\begin{equation}
{\dot P}={K\over P^3},~~~{\rm with}~K\equiv
{512\pi^4GI\epsilon_e^2\over 5c^5}\simeq {4096\pi^5G\over
75c^5}\rho\epsilon_e^2R_{\rm eff}^5,
\label{P}
\end{equation}
where the pulsar mass is approximated by $M\simeq 4\pi R_{\rm
eff}^3\rho/3$.

The solution of $P$ from Eq.(\ref{P}) is $P^4=P_i^4+4Kt$ ($P_i$
denotes the initial spin period), which is shown in Fig. 1 for
different parameters of $\epsilon_e$ and $R_{\rm eff}$
($\rho=4\times 10^{14}$ g/cm$^3$ is assumed).
\begin{figure}[ht]
\centerline{\epsfxsize=10cm\epsfbox{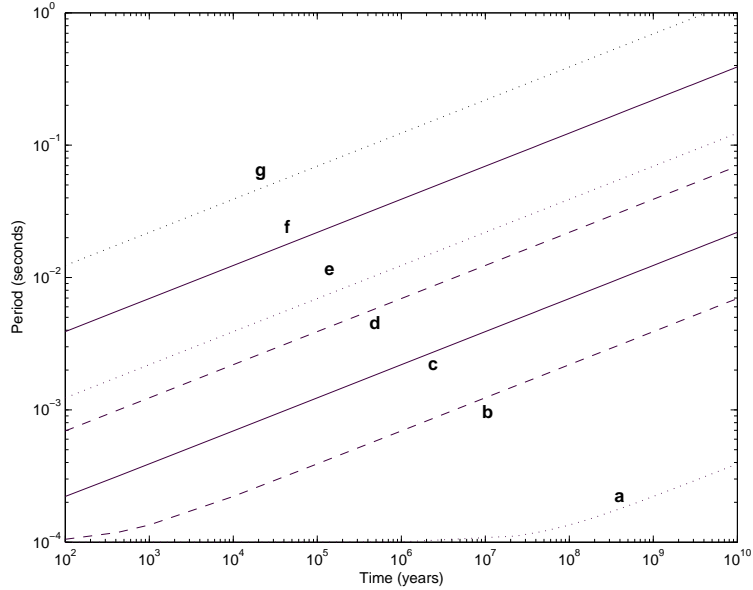}}
\caption{%
Gravitational-wave-radiation-induced period evolution of
sub-millisecond pulsars with an initial period $P_i=0.1$ ms for
different parameters of equatorial ellipticity, $\epsilon_e$, and
effective radius, $R_{\rm eff}$, but a fixed density $\rho=4\times
10^{14}$ g/cm$^3$. The lines are labelled from ``{\bf a}'' to
``{\bf g}''. ``a'': $\epsilon_e=10^{-3}$, $R_{\rm eff}=0.01$ km;
``b'': $\epsilon_e=10^{-3}$, $R_{\rm eff}=0.1$ km; ``c'':
$\epsilon_e=0.01$, $R_{\rm eff}=0.1$ km; ``d'': $\epsilon_e=0.1$,
$R_{\rm eff}=0.1$ km; ``e'': $\epsilon_e=10^{-3}$, $R_{\rm eff}=1$
km; ``f'': $\epsilon_e=0.01$, $R_{\rm eff}=1$ km; ``g'':
$\epsilon_e=0.1$, $R_{\rm eff}=1$ km.
\label{inter}}
\end{figure}
We see that the period increase rate is proportional to
$(\epsilon_e^{2/5}R_{\rm eff})^5$, a much small value of
$\epsilon_e$ and/or a low mass could make it possible that
sub-millisecond pulsars exist in the Universe.
Since $K\propto \rho R_{\rm eff}^5\sim MR_{\rm eff}^2$, $P$ would
increase much faster for normal neutron stars than that for quark
stars because of the sharp difference between the mass-radius
relations of these two kinds of stars, if the equatorial
ellipticities, $\epsilon_e$, are almost the same.
The origin of $\epsilon_e$ could be due to the secular instability
of Maclaurin spheroid (to become Jacobi ellipsoid finally) before
solidification, or the glitch-induced bumpy surface after
solidification.
Detailed calculation on $\epsilon_e$ is necessary and essential
for us to know the spindown features of sub-millisecond pulsars as
well as to estimate the gravitational wave strength emitted from
such stars.
It is worth noting that quark nuggets (stars) born during cosmic
QCD phase separation may exist as super-millisecond pulsars due to
spindown in Hubble time ($\sim 10^{10}$ years) although they might
still keep sub-millisecond spin if their values of $K$ is small
enough.

\subsection{Propeller-torqued spindown}

Besides spindown mechanisms due to gravitational wave radiation
and to magnetospheric activity, another one could also work when a
sub-millisecond pulsar is in a propeller phase.
Propeller torque acts on a pulsar through magnetohydrodynamical
(MHD) coupling at magnetospheric boundary with Alfv\'en radius,
and accreted matter inward has to been ejected at the boundary.
The matter going outward gains kinematic energy during this
process, and the pulsar losses then its rotation energy.

A real difficulty to estimate quantitatively propeller-torqued
spindown is to know the accretion rate, $\dot M$.
The accreted matter could be either the debris captured during the
birth of pulsars or the inter-stellar medium. Such accretion
details, including the MHD coupling, have not been known with
certainty yet.

\section{Magnetospheric activity of sub-millisecond pulsars}

The potential drop in the open-field-line region is essential for
the magnetospheric activity of sub-millisecond pulsars.
In case of approximately constant $\mu_{\rm m}$, the potential
drop between the center and the edge of a polar cap can be
expressed as\cite{xu05},
\begin{equation}
\phi = {64\pi^3\over 3 c^2}{\bar B}\mu_mR_{\rm eff}^3P^{-2}\simeq
2.2\times 10^{13}{\rm (volts)}~\mu_{m-6}R_{\rm effkm}^3P_{\rm
ms}^{-2},
\label{drop}
\end{equation}
where the bag constant ${\bar B}=60$ MeV/fm$^3\simeq 10^{14}$
g/cm$^3$ (i.e., $\rho/4$).
It is well known that pair production mechanism is a key
ingredient for pulsar radio emission. A pulsar is called to be
``death'' if the pair production condition can not be satisfied.
Although a real deathline depends upon the dynamics of detail pair
and photon production, the deathline can also be conventionally
taken as a line of constant potential drop $\phi$.
Assuming a critical drop $\phi_c=10^{12}$ volts, a sub-millisecond
pulsar with $P=0.1$ ms could still be active even its radius is
only 0.08 km, in case of $\mu_{m-6}=1$.

The potential drop in the open field line region would be much
higher than that presented in Eq.(\ref{drop}) if the effect of
inclination angle is included\cite{ycx06}.
Note that this conclusion favors the magnetospheric activity of
sub-millisecond pulsars.

Part of the power of the magnetospheric activity is in the
electro-magnetic emission of radio band.
If the radio power accounts for $\eta \approx 10^{-10\sim -5}$
times of the magnetospheric activity\cite{lk05}, the radio
luminosity is then, from Eq.(\ref{lossrate}),
\begin{equation}
L_{\rm radio}=\eta {512\pi^6\over 27c^3}\mu_m^2\rho^2R_{\rm
eff}^6P^{-4}\simeq 1.1\times 10^{32} \eta{(\rm
erg/s)}~\mu_{m-6}^2R_{\rm effkm}^6P_{\rm ms}^{-4},
\label{Lradio}
\end{equation}
where $\rho=4\times 10^{14}$ g/cm$^3$. This power is in the same
order of that of normal radio pulsars observed\cite{lk05} even the
radius of sub-millisecond pulsars is less than 1 km.
Although sub-millisecond pulsars could be radio loud, one needs a
very short sampling time, and has to deal with then a huge amount
of data in order to find a sub-millisecond pulsar.
Due to its large receiving area and wide scanning sky, the future
radio telescope, {\em FAST} (five hundred meter aperture spherical
telescope), to be built in Yunnan, China, might uncover
sub-millisecond radio pulsars.

\section{Conclusions and discussions}

We show that sub-millisecond pulsars should be in Jacobi
ellipsoidal figures of equilibrium.
It is addressed that the spindown of sub-millisecond pulsars would
be torqued dominantly by gravitational wave radiation, and that
such pulsars may not spin down to super-millisecond periods via
gravitation wave radiation during their lifetimes if they are
extremely low mass bare strange quark stars.
It is possible, based on the calculation of Fig. 1, that isolated
super-millisecond pulsars could be quark nuggets (stars) born
during cosmic QCD phase separation (via spindown in Hubble
timescale).

The radio luminosity of sub-millisecond pulsars could be high
enough to be recorded in advanced radio telescopes (e.g., the
future {\em FAST} in China).
Sub-millisecond pulsars would not be likely to be normal neutron
stars. It could then be a clear way of identifying quark stars as
the real nature of pulsars to search and detect sub-millisecond
radio pulsars.

Where to find sub-millisecond radio pulsars?
This is a question related to how sub-millisecond pulsars origin.
Actually, a similar question, which was listed as one of
Lorimer-Kramer's 13 open questions in pulsar astronomy\cite{lk05},
is still not answered: How are isolate millisecond pulsars
produced?
More further issues are related: Should sub-millisecond pulsars be
in globular clusters? Can millisecond and sub-millisecond pulsars
form during cosmic QCD separation? How to estimate an initial
period of quark star in this way? Could AIC (accretion-induced
collapse) of white dwarfs produce pulsars with periods $<10$ ms? A
recent multi-dimensional simulations of AIC was done\cite{aic06}
in the normal neutron stars regime, but a quark-star version of
AIC simulation is interesting and necessary.
Another interesting idea is: could low mass quark stars form
during the fission of a progenitor quark star? (Quark matter
produced in this way might be chromatically charges?). All these
ideas are certainly interesting, and could {\em not} be ruled out
simply and quickly by first principles.

\section*{Acknowledgments}
The author would like to thank many stimulating discussions with
the members in the pulsar group of Peking University.

\end{document}